\documentclass[conference]{IEEEtran}
\IEEEoverridecommandlockouts
\usepackage[T1]{fontenc}
\usepackage{microtype}
\usepackage{cite}
\usepackage{booktabs}
\usepackage{graphicx}
\usepackage{amsmath}
\usepackage{xspace}
\usepackage{tikz}
\usetikzlibrary{positioning,arrows.meta,fit}
\usepackage[hidelinks]{hyperref}

\newcommand{\devTargets}{100}
\newcommand{\devBuildable}{94}
\newcommand{\devMigrations}{5}
\newcommand{\testTargets}{123}
\newcommand{\testBuildable}{122}
\newcommand{\testMigrations}{7}
\newcommand{\devFrontier}{72}
\newcommand{\devSpinfer}{24}
\newcommand{\devQwenCoderOneShot}{21.3}

\newcommand{\devQwenCoderFull}{38.0}

\newcommand{\testFrontier}{66}
\newcommand{\testSpinfer}{41}
\newcommand{\testQwenCoderOneShot}{1.0}

\newcommand{\testQwenCoderFull}{35.3}
\newcommand{\testQwenCoderFullMin}{32}
\newcommand{\testQwenCoderFullMax}{41}

\newcommand{\testLlamaFull}{7.0}

\newcommand{\testLlamaFullMax}{12}

\newcommand{\testGemmaFull}{2.7}

\newcommand{\testAblQwenCoderOneShot}{1.0}

\newcommand{\devAblQwenCoderFeedback}{21.3}

\newcommand{\testAblQwenCoderFeedback}{1.0}

\newcommand{\devAblQwenCoderDecomp}{21.7}

\newcommand{\testAblQwenCoderDecomp}{18.0}

\newcommand{\devAblQwenCoderGuidance}{29.0}

\newcommand{\testAblQwenCoderGuidance}{35.3}
\newcommand{\testAblLlamaGuidance}{10.0}

\newcommand{\devAblQwenCoderRepair}{38.0}

\newcommand{\testAblLlamaRepair}{7.0}

\newcommand{\devSafeFrontierPrecision}{0.79}

\newcommand{\devSafeQwenCoderPrecision}{0.61}

\newcommand{\testSafeFrontierPrecision}{0.70}

\newcommand{\testSafeSpinferBroken}{13.0}
\newcommand{\testSafeSpinferPrecision}{0.55}
\newcommand{\testSafeQwenCoderPreDivergent}{22.0}
\newcommand{\testSafeQwenCoderPreBroken}{13.0}
\newcommand{\testSafeQwenCoderPrePrecision}{0.48}
\newcommand{\testSafeQwenCoderDivergent}{8.7}
\newcommand{\testSafeQwenCoderBroken}{0.0}
\newcommand{\testSafeQwenCoderPrecision}{0.74}

\newcommand{\parseFailQwenCoderNoFix}{44}

\newcommand{\parseFailQwenCoderFix}{29}

\newcommand{\parseFailLlamaNoFix}{43}

\newcommand{\parseFailLlamaFix}{38}

\newcommand{\parseFailGemmaNoFix}{100}

\newcommand{\parseFailGemmaFix}{42}

\newcommand{\testCostQwenCoderCalls}{9.7}
\newcommand{\testCostQwenCoderSeconds}{15}

\begin{document}

\title{Guided Generation of Linux Kernel Semantic Patches with Small Local Language Models}

\author{\IEEEauthorblockN{Mohanned Ahmed}
\IEEEauthorblockA{ECCE School, Faculty of Engineering, Egypt-Japan University of Science and Technology\\
Alexandria, Egypt\\
Email: mohanned.hafez@ejust.edu.eg}}

\maketitle

\begin{abstract}
When a Linux kernel API changes, hundreds of drivers are converted by hand over months, and the earlier
conversions are reviewed examples for the later ones. We ask whether small open language models running on a
laptop can turn such examples into Coccinelle semantic patches, the rules kernel maintainers already use for
tree-wide changes. We adapt guided code generation to this task: the training examples are decomposed by script
into a tree of change variants, the model writes one small rule per variant under feedback from applying it to the
examples, deterministic components repair SmPL syntax and generalise driver-specific names, rules are composed
greedily, and a compile critic repairs or rejects a rule whose output does not build. We evaluate on a
pre-registered test set of \testMigrations{} migrations whose new APIs appeared after the training cutoffs of all
three models (Qwen2.5-Coder 7B, Llama~3.1 8B, Gemma~3 4B), with \testBuildable{} held-out driver conversions,
counting a conversion as solved only if the rule's output matches the human fix and compiles. With the framework,
Qwen2.5-Coder solves \testQwenCoderFull{} conversions on average over three seeds (range
\testQwenCoderFullMin--\testQwenCoderFullMax), against \testQwenCoderOneShot{} when prompted directly; Spinfer
solves \testSpinfer{} and a frontier LLM agent \testFrontier. The local model solves fewer conversions than
Spinfer but its rules are the most precise on the test set (\testSafeQwenCoderPrecision{} of the files they change
are solved, against \testSafeSpinferPrecision{} for Spinfer and \testSafeFrontierPrecision{} for the frontier
agent), and the compile critic removes every output that matches the human fix but does not build. The two general-purpose
models stay weak even with guidance (at most \testLlamaFullMax{} solved). All code, prompts, model responses and
results are released.
\end{abstract}


\section{Introduction}
\label{sec:intro}

When a Linux kernel API changes, the drivers that use it are converted one after another, often over months and by
different developers. The earlier conversions are free, reviewed examples for the later ones: in a pilot study of
27 API migrations in Linux v5.10--v6.10, an earlier conversion of another file by the same migration (a
\emph{sibling migration}) existed in the target's git history for 96.9\% of 8{,}818 conversions~\cite{ahmed2026siblings}.
The natural way to reuse such examples in the kernel is a Coccinelle semantic patch~\cite{padioleau2008documenting}:
a rule that kernel maintainers can read, review and apply tree-wide, unlike a stream of per-file edits from a
language model.

Writing such rules automatically from examples is an old problem. Spinfer~\cite{serrano2020spinfer} infers them
with dedicated heuristics, and recent work prompts frontier LLMs to write them~\cite{yusuf2025howfar}, finding
that LLM-written patches are often syntactically invalid and that both approaches struggle when the examples
contain several variants of a change. Frontier APIs are also a poor fit for kernel workflows: maintainers cannot
inspect them, and they have likely seen the very migrations they are evaluated on.

We ask whether \emph{small open models running on a laptop} can write correct migration rules when the work is
structured for them. We adapt guided code generation (GCG)~\cite{almorsi2025gcg}, a decomposition framework
previously shown to help an 8B model on HumanEval, to semantic patches: the training siblings are decomposed into
a tree of change variants; the model writes one small rule per variant under test feedback from the siblings
themselves; deterministic components repair mechanical SmPL errors and generalise driver-specific names; rules are
composed greedily; and a compile critic either repairs or rejects a rule whose output does not build.

On \testMigrations{} migrations whose APIs postdate all three models, the framework takes Qwen2.5-Coder 7B from
\testQwenCoderOneShot{} to \testQwenCoderFull{} solved conversions out of \testBuildable{}, close to but below
Spinfer (\testSpinfer), with the highest precision of all systems on the test set
(\testSafeQwenCoderPrecision). Decomposition and guidance carry most of the gain; the compile critic adds no
solved conversions but removes every output that matches the human fix and fails to build. The two general-purpose
models, Llama~3.1 8B and Gemma~3 4B, remain weak, mainly because they rarely write SmPL that parses. Changes that
span a declaration and its uses, such as a callback whose return type changes, defeat every local model.

\paragraph{Contributions.}
\begin{itemize}
  \item A guided framework for semantic-patch generation with 4--8B local models (Section~\ref{sec:method}), whose
    every step except rule writing is deterministic and auditable.
  \item A pre-registered test set beyond the models' training cutoffs: \testMigrations{} Linux API migrations whose new API first
    appeared after all three models' training cutoffs, with \testTargets{} held-out driver conversions
    (Section~\ref{sec:setup}).
  \item An evaluation that compiles candidate outputs on x86\_64 \texttt{allmodconfig} and reports an automatic,
    judgement-free success metric and precision, three seeds per configuration, an ablation of every component,
    and an analysis of the failures (Section~\ref{sec:results}).
\end{itemize}
All code, prompts, model responses and results are released.\footnote{\url{https://github.com/MohannedAhmed67/kernel-sibling-migrations}}

\section{Related Work}
\label{sec:related}

\paragraph{Collateral evolutions and semantic patches.} Padioleau et al.\ named the driver changes forced by kernel
API changes \emph{collateral evolutions}~\cite{padioleau2006understanding} and introduced Coccinelle and its
semantic patch language SmPL to perform them~\cite{padioleau2008documenting}. Such rules are usually written by
experts. Our training data are earlier hand conversions of the same migration (\emph{sibling migrations}).

\paragraph{Inferring rules from examples.} spdiff infers a generic patch from a few before/after file
pairs~\cite{andersen2008generic}. Spinfer~\cite{serrano2020spinfer} infers Coccinelle rules from example changes and
already evaluates chronologically, learning from the first changed files and testing on later ones; its evaluation
compares inferred rules with expert patches rather than compiling the outputs, and in our runs it inferred no rule
for migrations whose change lies in a struct initialiser outside any function (``no code to learn from'').
Prequel retrieves relevant porting commits from history~\cite{lawall2017prequel}, and FixMorph synthesises a
transformation from one patch for backporting~\cite{shariffdeen2021fixmorph}.

\paragraph{LLMs and semantic patches.} Yusuf et al.\ compare Spinfer with GPT-4, Claude~3.7 Sonnet and DeepSeek-V3
prompted to write SmPL from change examples~\cite{yusuf2025howfar}. They find that the LLMs produce valid patches
from few examples but struggle with syntax, that Spinfer produces overly specific patches, and that both perform
poorly on change examples with several variants; among their future directions they list refining abstraction
``based on feedback from patch application results''. We build exactly such a feedback loop, add decomposition by
variant and a compile critic, use 4--8B models that run on a laptop instead of frontier APIs, and evaluate on a
pre-registered test set whose APIs postdate the models.

\paragraph{LLMs for kernel evolution.} AutoDriver (DRIVEBENCH)~\cite{kharlamova2025autodriver} updates drivers with
an LLM multi-agent loop checked by compilation and QEMU, without retrieving earlier conversions; MigGPT~\cite{dang2025miggpt}
and PortGPT~\cite{li2026portgpt} migrate or backport out-of-tree patches with LLMs. These systems edit code directly; we
generate a reusable, reviewable rule that kernel maintainers already know how to read and apply tree-wide.

\paragraph{Decomposition and critics for code generation.} Our framework adapts guided code generation
(GCG)~\cite{almorsi2025gcg}, which decomposes a task into a tree, generates bottom-up with tests and adds critic
agents, and which improved Llama~3.1 8B on HumanEval. GCG reported no ablations or variance; here each component
is ablated over three seeds and three models.

\section{Guided Semantic-Patch Generation}
\label{sec:method}

\begin{figure*}[t]
\centering
\begin{tikzpicture}[
  node distance=5mm and 5mm,
  box/.style={draw, rounded corners, align=center, font=\footnotesize, minimum height=12mm, text width=30mm},
  det/.style={box, fill=gray!12},
  llm/.style={box, fill=blue!10},
  arr/.style={-{Stealth[length=2mm]}}]
\node[det] (s1) {\textbf{1.~Variant tree}\\hunks $\to$ leaves\\(script)};
\node[llm, right=of s1] (s2) {\textbf{2.~Leaf rules}\\model + \texttt{spatch}\\feedback, $\le 3$ rounds};
\node[det, right=of s2] (s3) {\textbf{3.~Composition}\\greedy, scored on\\all siblings (script)};
\node[llm, right=of s3] (s4) {\textbf{4.~Compile critic}\\build, repair\\or abstain};
\node[det, below=of s2, text width=62mm, minimum height=7mm] (fix)
  {auto-repair $\cdot$ generality critic $\cdot$ lint (script)};
\node[above=2mm of s1, font=\footnotesize] (in) {training siblings};
\node[above=2mm of s4, font=\footnotesize] (out) {one rule, or abstain};
\draw[arr] (in) -- (s1); \draw[arr] (s1) -- (s2); \draw[arr] (s2) -- (s3); \draw[arr] (s3) -- (s4);
\draw[arr] (s4) -- (out);
\draw[arr] ([xshift=-3mm]s2.south) -- ([xshift=-3mm]s2.south |- fix.north);
\draw[arr] ([xshift=3mm]s2.south |- fix.north) -- ([xshift=3mm]s2.south);
\end{tikzpicture}
\caption{The four stages. Shaded boxes are deterministic scripts; blue boxes call the local model. Only the
training siblings are read; held-out targets are never seen.}
\label{fig:pipeline}
\end{figure*}
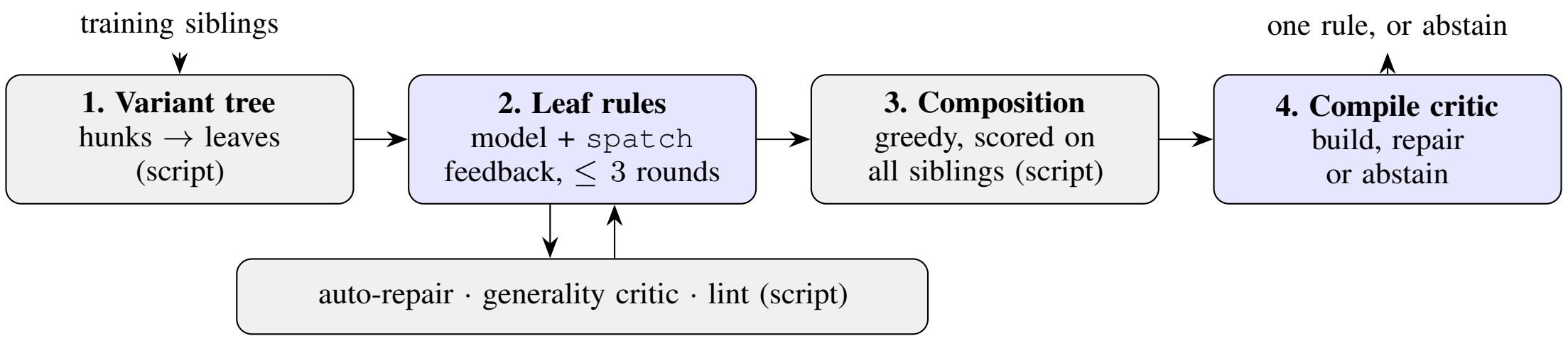

The input is a migration $M$ (its id and the old/new API line patterns used to mine it) and its training siblings:
pairs $(b_i, a_i)$ of one driver file before and after its hand conversion, all committed before any target was
written. The output is one Coccinelle semantic patch for $M$, or an explicit abstention. No step reads a held-out
target. The design follows guided code generation (GCG)~\cite{almorsi2025gcg}: decompose the task into a tree,
generate the leaves bottom-up with tests, and let a critic check the assembled result. Here the ``tests'' are the
training siblings themselves, and the critic is the C compiler. Figure~\ref{fig:pipeline} shows the four stages.

\paragraph{Units and change hunks.} Both versions of a file are split into statement-level units (cut at
\texttt{;}, \texttt{\{}, \texttt{\}} and comment ends; preprocessor lines are units of their own) and normalised for
whitespace. A hunk is a maximal run of removed and added units. All scoring below counts units: a rule's output
$o_i$ on $b_i$ \emph{hits} an expected unit if the unit change also appears between $b_i$ and $o_i$, and makes an
\emph{extra} edit for every unit change between $b_i$ and $o_i$ that the human did not make between $b_i$ and $a_i$.

\subsection{Stage 1: variant tree (decomposition)}
Every hunk of every training sibling becomes a leaf candidate. Hunks that touch a migration API name (the first
identifier of a match of the old or new pattern) are \emph{anchor} hunks and are grouped by the API names they
contain; the remaining hunks are \emph{companion} hunks (for example, a function's return type that changes because
the API it is registered with changed) and are grouped by their C keywords and by identifiers that recur in at least
two training files. Each group is a leaf. Within a leaf, hunks are abstracted to a \emph{shape}: identifiers become
\texttt{ID1}, \texttt{ID2},~\ldots, numbers \texttt{NUM} and strings \texttt{STR}; the \emph{minimal edit} is the
hunk with its common prefix and suffix tokens removed. A leaf is a \emph{task} if it is an anchor leaf or occurs in
at least two training files; companion leaves seen in one file only are treated as possibly unrelated edits.

\subsection{Stage 2: leaf rules with test feedback}
For each task leaf, the model receives a short SmPL primer (a toy \texttt{kmalloc}+\texttt{memset}~$\to$
\texttt{kzalloc} rule and syntax notes, unrelated to any studied migration) and a task prompt with one example per
distinct minimal-edit shape: the enclosing function or struct header (\emph{inside}), the preceding context, the
minimal edit and its abstract pattern. Anchor leaves are generated first. A companion leaf gets an
\emph{interface}: the framework takes the passing rule of the anchor leaf that shares most training files with it,
names its first rule, and writes the header of a dependent rule that inherits all its metavariables; the model only
writes the body. This is GCG's ``a parent sees only its children's interfaces'', inverted for rules that must run
after their anchor.

Each answer goes through four deterministic steps before \texttt{spatch} runs:
(i) \emph{auto-repair} of mechanical SmPL errors (missing or trailing \texttt{@@}, metavariables used as fields or
callees declared as \texttt{identifier}, undeclared \texttt{ID}/\texttt{NUM}/\texttt{STR} placeholders);
(ii) the \emph{generality critic}: a name that the rule matches literally, that the change keeps (it appears on
context lines or on both a removed and an added line), that is not called as a function, and that occurs in only
one training file is driver-specific and is turned into a metavariable;
(iii) a \emph{lint} for rules without \texttt{-}/\texttt{+} lines, comments used in place of code, and unbalanced
brackets; (iv) application to every training file of the leaf. If the rule fails to parse, misses an expected
unit, or makes an extra edit, the model gets feedback for up to $R=3$ rounds: \texttt{spatch}'s error with a
matching generic repair hint, or, per failing file, the missed units with their abstract pattern and the extra edits.
A rule identical to an earlier one raises the sampling temperature by 0.3 for the next round. The best round is
kept (parses, then most hits, then fewest extra edits, then earliest).

\subsection{Stage 3: composition}
Leaf rules are renamed so that rule names cannot collide (a companion rule carries its anchor rule with it) and
composed greedily: candidates are tried in decreasing order of their own hits minus extra edits, and a candidate is
kept only if the composed patch still parses and its score over \emph{all} training files of the migration
strictly increases. Expected units are those of all task leaves, so a leaf that also helps another leaf is
rewarded and a leaf whose edits clash with another is dropped.

\subsection{Stage 4: compile critic}
The composed rule is applied to the most recent training files it changes (up to three whose human version
compiles), and each output is compiled in a worktree at that training commit's parent (x86\_64
\texttt{allmodconfig}, \texttt{CONFIG\_WERROR} off). If an output does not compile, the model gets the compiler
errors, the converted code around them, and the diff between its output and the human version of that training
file, and writes one additional dependent rule through the same interface mechanism (two rounds). Among the rules
whose checked outputs compile (the composed rule, the repaired rules, and the composed rule minus one leaf), the
one with the best training score is kept; if none compiles, the framework \emph{abstains}. A migration without a
rule is counted as a failure on every target, never silently dropped.

\section{Experimental Setup}
\label{sec:setup}

\begin{table*}[t]
\centering
\caption{Data. \emph{New API}: first commit that adds the new API under \texttt{include/} (test set only; all
after the models' training cutoffs). \emph{Train}: training siblings. \emph{Buildable}: targets whose original file
compiles on x86\_64 \texttt{allmodconfig}; the denominator of every success count.}
\label{tab:data}
\footnotesize
\begin{tabular}{llrrrrl}
\toprule
Set & Migration & New API & Train & Targets & Buildable & Target dates \\
\midrule
dev & \texttt{ida\_simple} & -- & 20 & 20 & 20 & 2022-06--2024-05 \\
dev & \texttt{i2c\_probe\_new} & -- & 20 & 20 & 19 & 2023-05--2023-09 \\
dev & \texttt{platform\_remove\_new} & -- & 20 & 20 & 20 & 2023-03--2024-05 \\
dev & \texttt{strlcpy\_strscpy} & -- & 20 & 20 & 15 & 2021-01--2023-12 \\
dev & \texttt{pci\_dma\_wrappers} & -- & 20 & 20 & 20 & 2020-12--2022-02 \\
\midrule
test & \texttt{secs\_to\_jiffies} & 2024-10-30 & 17 & 10 & 10 & 2025-04--2026-05 \\
test & \texttt{iio\_claim\_direct} & 2025-02-17 & 20 & 20 & 20 & 2025-04--2025-04 \\
test & \texttt{iio\_push\_with\_ts} & 2025-05-21 & 20 & 20 & 20 & 2025-08--2026-08 \\
test & \texttt{gpio\_set\_rv} & 2025-02-26 & 20 & 20 & 20 & 2025-04--2025-07 \\
test & \texttt{pcim\_request\_all\_regions} & 2024-10-30 & 18 & 17 & 16 & 2025-04--2026-08 \\
test & \texttt{ethtool\_rx\_ring\_count} & 2025-09-18 & 15 & 20 & 20 & 2025-11--2026-01 \\
test & \texttt{devm\_kmemdup\_array} & 2025-02-24 & 7 & 16 & 16 & 2025-02--2025-12 \\
\bottomrule
\end{tabular}

\end{table*}

\begin{table*}[t]
\centering
\caption{Test set: held-out conversions solved (output matches the human fix exactly or as a complete subset, and
compiles). Local models: mean over three seeds, [min--max] where the seeds differ. \textsc{one-shot}: one rule from
the plain examples; \textsc{full}: the complete framework. $^\dagger$A frontier LLM agent that may have seen these
migrations during training; a reference point, not a like-for-like competitor.}
\label{tab:test}
\footnotesize
\setlength{\tabcolsep}{4pt}
\begin{tabular}{lrrrrrrrrr}
\toprule
 & & Frontier & & \multicolumn{2}{c}{Qwen2.5-Coder 7B} & \multicolumn{2}{c}{Llama 3.1 8B} & \multicolumn{2}{c}{Gemma 3 4B} \\
\cmidrule(lr){5-6}\cmidrule(lr){7-8}\cmidrule(lr){9-10}
Migration & Bld. & LLM$^\dagger$ & Spinfer & one-shot & \textsc{full} & one-shot & \textsc{full} & one-shot & \textsc{full} \\
\midrule
\texttt{devm\_kmemdup\_array} & 16 & 1 & 2 & 0.0 & 1.3 [0--2] & 0.0 & 0.0 & 0.0 & 0.0 \\
\texttt{ethtool\_rx\_ring\_count} & 20 & 12 & 0 & 0.0 & 0.0 & 0.0 & 0.0 & 0.0 & 0.0 \\
\texttt{gpio\_set\_rv} & 20 & 14 & 0 & 0.0 & 0.0 & 0.0 & 0.0 & 0.0 & 0.0 \\
\texttt{iio\_claim\_direct} & 20 & 17 & 16 & 0.0 & 16.0 & 0.0 & 0.0 & 0.0 & 0.0 \\
\texttt{iio\_push\_with\_ts} & 20 & 12 & 12 & 1.0 & 0.7 [0--1] & 0.0 & 0.0 & 0.0 & 0.0 \\
\texttt{pcim\_request\_all\_regions} & 16 & 8 & 11 & 0.0 & 15.0 & 0.0 & 7.0 [0--12] & 0.0 & 2.7 [0--8] \\
\texttt{secs\_to\_jiffies} & 10 & 2 & 0 & 0.0 & 2.3 [0--7] & 0.0 & 0.0 & 0.0 & 0.0 \\
\midrule
\textbf{All} & 122 & 66 & 41 & 1.0 & 35.3 [32--41] & 0.0 & 7.0 [0--12] & 0.0 & 2.7 [0--8] \\
\bottomrule
\end{tabular}

\end{table*}

\paragraph{Migrations and split.} A \emph{conversion} is a (commit, file) pair whose diff removes a line matching the
migration's old-API pattern and adds a line matching its new-API pattern. Per migration, the earliest 30\% of
\texttt{.c} conversions by commit time (at most 20, one per file) are the training siblings. Targets are conversions
\emph{written} (author time) after the last training commit was committed, in a file and a commit that are not
training ones; up to 20 are sampled evenly over time. Every training sibling is therefore in a different file and
existed before the target was written.

\paragraph{Development set.} \devMigrations{} migrations from Linux v5.10--v6.10 with \devTargets{} targets
(Table~\ref{tab:data}). All prompts, hints, thresholds and critic rules were designed on their training siblings, and their
held-out targets were evaluated during development, so results on them are development results.

\paragraph{Test set (pre-registered).} Before mining any history after v6.10, we fixed 14 candidate migrations whose
new API we expected to be recent, and an inclusion rule (committed to the repository before the mined data existed):
(1) the new API is absent from \texttt{include/} at v6.10 and first appears on or after 2024-09-01, after the
training cutoff of all three models; (2) at least three conversion commits; (3) verified precision of at least 0.85
on up to eight sampled candidates, each judged by two independent LLM reviewers against the real diff; (4) at least
five training siblings and ten eligible targets after the split. Seven migrations qualified
(Table~\ref{tab:data}); three of the excluded ones fail rule~4 because almost all of their conversions were
\emph{written} before the last training commit (a patch series posted at once and merged over months). The test set
was run once, with the framework frozen at a git tag.

\paragraph{Models.} Three open-weight models run locally with Ollama on one 8\,GB laptop GPU (RTX~5060, 4-bit
quantisation): Qwen2.5-Coder 7B~\cite{qwen25coder} (code data from repositories created before February 2024), Llama~3.1 8B~\cite{llama3}
(knowledge cutoff December 2023) and Gemma~3 4B~\cite{gemma3} (August 2024). Context 8{,}192 tokens, temperature 0.4, at most
1{,}024 output tokens, seeds 1--3.

\paragraph{Baselines.} \emph{Spinfer}~\cite{serrano2020spinfer} (commit \texttt{c47696a}, default settings) on the
same training pairs. \emph{Frontier LLM}: a Claude agent (Anthropic) whose file access was limited to the
migration's training folder writes one rule per migration and may test it with \texttt{spatch} on the training files;
a transcript audit checks that it read nothing else. Unlike the local models, this agent's training data may contain
the test migrations, so it is an upper reference, not a fair competitor.

\paragraph{Ablations.} Six configurations, each adding one component: \textsc{one-shot} (one rule for the whole
migration from the plain examples, no feedback), \textsc{+feedback} (three rounds), \textsc{+decomposition} (leaf
tasks, plain prompts), \textsc{+guidance} (minimal-edit patterns, context, interface, hints, lint, stall handling),
\textsc{+auto-repair}, and \textsc{full} (+generality critic and compile critic). All configurations are composed by
stage~3.

\paragraph{Evaluation.} Each rule is applied with \texttt{spatch}~1.1.1 to the target file at the parent of its fix
commit and compared with the human fix on whitespace-normalised lines: \emph{exact} (identical), \emph{subset
complete} (every changed line is one the human changed, and no more old-API uses remain than in the human
version), \emph{subset incomplete}, \emph{divergent} (an edit the human did not make), \emph{no change},
\emph{spatch error} and \emph{no rule}. The original, human and rule versions are compiled in that tree (x86\_64
\texttt{allmodconfig}, \texttt{CONFIG\_WERROR} off). A target is \emph{buildable} if its original file compiles.
\textbf{Success} means exact or subset complete \emph{and} the rule output compiles, counted over buildable targets.
This metric needs no human judgement; it undercounts correct rules that differ from the human fix in form
(Section~\ref{sec:threats}). We also report \emph{precision}, the share of targets a rule edits that it solves,
since a rule that edits a file wrongly costs a maintainer more than a rule that leaves it alone.

\section{Results}
\label{sec:results}

\paragraph{Main result.} Table~\ref{tab:test} gives the test-set results. Prompted directly, none of the three local
models can turn the training siblings into a working rule: Qwen2.5-Coder solves \testQwenCoderOneShot{} of
\testBuildable{} conversions and the other two none. With the framework, Qwen2.5-Coder solves
\testQwenCoderFull{} on average (\testQwenCoderFullMin--\testQwenCoderFullMax{} across seeds), Llama~3.1 8B
solves \testLlamaFull{} and Gemma~3 4B \testGemmaFull. Spinfer solves \testSpinfer{} and the frontier LLM agent
\testFrontier. The local model therefore closes most but not all of the distance to Spinfer, and the frontier agent
stays well ahead. The per-migration picture is uneven: the guided Qwen2.5-Coder solves more conversions than
Spinfer on \texttt{pcim\_request\_all\_regions} and \texttt{secs\_to\_jiffies}, matches it on
\texttt{iio\_claim\_direct}, and falls behind on \texttt{iio\_push\_with\_ts}; on \texttt{gpio\_set\_rv} and
\texttt{ethtool\_rx\_ring\_count} neither it nor Spinfer solves anything, while the frontier agent does.

On the development set (Table~\ref{tab:dev}), whose migrations shaped the design, Qwen2.5-Coder solves
\devQwenCoderFull{} of \devBuildable{} with the framework against \devQwenCoderOneShot{} one-shot, Spinfer
\devSpinfer{} and the frontier agent \devFrontier. The general-purpose models again stay near zero.

\begin{table*}[t]
\centering
\caption{Development set, same layout as Table~\ref{tab:test}. These migrations were used to design the framework,
so the numbers are optimistic.}
\label{tab:dev}
\footnotesize
\setlength{\tabcolsep}{4pt}
\begin{tabular}{lrrrrrrrrr}
\toprule
 & & Frontier & & \multicolumn{2}{c}{Qwen2.5-Coder 7B} & \multicolumn{2}{c}{Llama 3.1 8B} & \multicolumn{2}{c}{Gemma 3 4B} \\
\cmidrule(lr){5-6}\cmidrule(lr){7-8}\cmidrule(lr){9-10}
Migration & Bld. & LLM$^\dagger$ & Spinfer & one-shot & \textsc{full} & one-shot & \textsc{full} & one-shot & \textsc{full} \\
\midrule
\texttt{i2c\_probe\_new} & 19 & 19 & 0 & 12.7 [0--19] & 19.0 & 0.0 & 0.0 & 0.0 & 0.0 \\
\texttt{ida\_simple} & 20 & 14 & 9 & 0.0 & 5.3 [5--6] & 0.0 & 1.7 [0--5] & 0.0 & 3.0 [0--9] \\
\texttt{pci\_dma\_wrappers} & 20 & 6 & 2 & 0.0 & 1.3 [0--2] & 0.0 & 0.0 & 0.0 & 0.0 \\
\texttt{platform\_remove\_new} & 20 & 19 & 0 & 0.0 & 0.0 & 0.0 & 0.0 & 0.0 & 0.0 \\
\texttt{strlcpy\_strscpy} & 15 & 14 & 13 & 8.7 [0--13] & 12.3 [11--13] & 0.0 & 0.0 & 0.0 & 0.0 \\
\midrule
\textbf{All} & 94 & 72 & 24 & 21.3 [13--32] & 38.0 [37--39] & 0.0 & 1.7 [0--5] & 0.0 & 3.0 [0--9] \\
\bottomrule
\end{tabular}

\end{table*}

\begin{table*}[t]
\centering
\caption{Ablation: conversions solved as components are added (mean over three seeds, [min--max]). Every
configuration is composed by stage~3; the last row adds the compile critic of stage~4.}
\label{tab:ablation}
\footnotesize
\setlength{\tabcolsep}{4pt}
\begin{tabular}{lrrrrrr}
\toprule
 & \multicolumn{3}{c}{Development (of 94)} & \multicolumn{3}{c}{Test (of 122)} \\
\cmidrule(lr){2-4}\cmidrule(lr){5-7}
Configuration & Qwen2.5-Coder & Llama & Gemma & Qwen2.5-Coder & Llama & Gemma \\
\midrule
\textsc{one-shot} & 21.3 [13--32] & 0.0 & 0.0 & 1.0 & 0.0 & 0.0 \\
+ feedback & 21.3 [13--32] & 0.0 & 0.0 & 1.0 & 2.0 [0--6] & 0.0 \\
+ decomposition & 21.7 [13--32] & 0.0 & 0.0 & 18.0 & 2.0 [0--6] & 0.0 \\
+ guidance & 29.0 [21--34] & 2.3 [0--7] & 0.0 & 35.3 [32--41] & 10.0 [9--12] & 0.0 \\
+ auto-repair & 38.0 [37--39] & 1.7 [0--5] & 3.0 [0--9] & 35.3 [32--41] & 7.0 [0--12] & 2.7 [0--8] \\
+ generality critic & 38.0 [37--39] & 1.7 [0--5] & 3.0 [0--9] & 35.3 [32--41] & 7.0 [0--12] & 2.7 [0--8] \\
+ compile critic (\textsc{full}) & 38.0 [37--39] & 1.7 [0--5] & 3.0 [0--9] & 35.3 [32--41] & 7.0 [0--12] & 2.7 [0--8] \\
\bottomrule
\end{tabular}

\end{table*}

\paragraph{Which components matter.} Table~\ref{tab:ablation} adds one component at a time. For Qwen2.5-Coder on
the test set, test feedback alone changes nothing (\testAblQwenCoderOneShot{} to \testAblQwenCoderFeedback):
asked for one rule covering every variant at once, the model never produces a rule good enough for feedback to
refine. Decomposition into variant leaves is the first large step (\testAblQwenCoderDecomp), and guidance (minimal
edits, context, interfaces, hints, lint and stall handling) roughly doubles it again
(\testAblQwenCoderGuidance). On the development set the order differs: decomposition adds little
(\devAblQwenCoderFeedback{} to \devAblQwenCoderDecomp), guidance and auto-repair add most
(\devAblQwenCoderGuidance, then \devAblQwenCoderRepair). The general-purpose models fail for a more basic reason:
they rarely write SmPL that parses. Without auto-repair, the best rule of \parseFailGemmaNoFix\% of Gemma's leaf
tasks does not parse (\parseFailLlamaNoFix\% for Llama, \parseFailQwenCoderNoFix\% for Qwen2.5-Coder); auto-repair
lowers this to \parseFailGemmaFix\%, \parseFailLlamaFix\% and \parseFailQwenCoderFix\%. For Gemma, auto-repair
is what makes any success possible at all; for Llama on the test set it does not help (\testAblLlamaGuidance{}
before, \testAblLlamaRepair{} after). Adding the generality critic left the mean number of solved conversions
unchanged for every model on both sets. We keep it because it is cheap and did not hurt, but our data give no
evidence that it is needed.

\begin{table*}[t]
\centering
\caption{Precision of the outputs on buildable targets (local models: mean over seeds). \emph{Changed}: targets
the rule edits; \emph{Divergent}: the rule makes an edit the human did not make; \emph{Broken}: the output matches
the human fix (exactly or as a subset) but does not compile; \emph{Precision}: solved / changed.}
\label{tab:safety}
\footnotesize
\setlength{\tabcolsep}{5pt}
\begin{tabular}{llrrrrr}
\toprule
Set & Source & Success & Changed & Divergent & Broken & Precision \\
\midrule
dev & Frontier LLM$^\dagger$ & 72.0 & 91.0 & 17.0 & 0.0 & 0.79 \\
dev & Spinfer & 24.0 & 47.0 & 13.0 & 0.0 & 0.51 \\
dev & Qwen2.5-Coder 7B, before compile critic & 38.0 & 82.7 & 12.0 & 20.0 & 0.46 \\
dev & Qwen2.5-Coder 7B, \textsc{full} & 38.0 & 62.7 & 12.0 & 0.0 & 0.61 \\
dev & Llama 3.1 8B, before compile critic & 1.7 & 22.3 & 10.7 & 0.0 & 0.07 \\
dev & Llama 3.1 8B, \textsc{full} & 1.7 & 20.7 & 8.7 & 0.0 & 0.08 \\
dev & Gemma 3 4B, before compile critic & 3.0 & 9.3 & 2.7 & 0.0 & 0.32 \\
dev & Gemma 3 4B, \textsc{full} & 3.0 & 9.3 & 2.7 & 0.0 & 0.32 \\
\midrule
test & Frontier LLM$^\dagger$ & 66.0 & 94.0 & 28.0 & 0.0 & 0.70 \\
test & Spinfer & 41.0 & 74.0 & 18.0 & 13.0 & 0.55 \\
test & Qwen2.5-Coder 7B, before compile critic & 35.3 & 74.3 & 22.0 & 13.0 & 0.48 \\
test & Qwen2.5-Coder 7B, \textsc{full} & 35.3 & 48.0 & 8.7 & 0.0 & 0.74 \\
test & Llama 3.1 8B, before compile critic & 7.0 & 27.3 & 0.3 & 0.0 & 0.26 \\
test & Llama 3.1 8B, \textsc{full} & 7.0 & 27.3 & 0.3 & 0.0 & 0.26 \\
test & Gemma 3 4B, before compile critic & 2.7 & 34.7 & 0.0 & 12.0 & 0.08 \\
test & Gemma 3 4B, \textsc{full} & 2.7 & 22.7 & 0.0 & 0.0 & 0.12 \\
\bottomrule
\end{tabular}

\end{table*}

\paragraph{The compile critic buys precision at no cost.} The compile critic never changes the number of
solved conversions (Table~\ref{tab:ablation}), but it changes what the rule does to the others
(Table~\ref{tab:safety}). Before it, Qwen2.5-Coder's composed rules on the test set produce
\testSafeQwenCoderPreBroken{} outputs per seed that match the human fix but do not build, and
\testSafeQwenCoderPreDivergent{} divergent ones. The critic either repairs such a rule or abstains for that
migration, leaving \testSafeQwenCoderBroken{} broken and \testSafeQwenCoderDivergent{} divergent outputs and raising
precision from \testSafeQwenCoderPrePrecision{} to \testSafeQwenCoderPrecision{}, above Spinfer
(\testSafeSpinferPrecision, with \testSafeSpinferBroken{} broken outputs) and the frontier agent
(\testSafeFrontierPrecision). On the development set the frontier agent remains more precise
(\devSafeFrontierPrecision{} against \devSafeQwenCoderPrecision). For a maintainer, a rule that converts fewer
files but rarely needs to be undone, or no rule at all, is easier to adopt than one that has to be checked file by
file.

\paragraph{Where the framework fails.} Three failure modes account for the test migrations the local models do not
solve. (i) \emph{Changes that span declarations.} \texttt{gpio\_set\_rv} turns a \texttt{void} callback into one
returning \texttt{int}, returns the value of its last call, and renames the struct field that registers it;
\texttt{ethtool\_rx\_ring\_count} moves a \texttt{switch} case into a new function and registers it in an
\texttt{ethtool\_ops} initialiser. Both need several rules that agree on a function name across a file. The
framework's interface mechanism passes metavariables from an anchor rule to its dependants, but whenever a model
composed a rule for these migrations, its outputs on the training files did not compile (the compiler reports
the registering field renamed while the callback keeps its old type, or a \texttt{return} with a value left in a
\texttt{void} function), repair did not fix them, and the critic abstained
for all three models on every seed. The development
migration \texttt{platform\_remove\_new} has the same shape and the same outcome. (ii) \emph{Over-specific
patterns.} On \texttt{iio\_push\_with\_ts} Qwen2.5-Coder's rule requires the timestamp argument to be a function
call, so it misses the common \texttt{pf->timestamp} form, and it writes \texttt{sizeof(ID2)} where
\texttt{ID2} is the expression \texttt{\&scan}: the output computes the size of a pointer. That output compiles,
which no compile check can catch; it shows up only as a divergent edit. (iii) \emph{Invalid SmPL} remains the main
failure of the two general-purpose models, as quantified above.

\begin{table}[t]
\centering
\caption{Cost per migration of the \textsc{full} configuration on one laptop GPU (RTX~5060, 8\,GB): model calls
and seconds of generation in stage~2, and repair calls of the compile critic (stage~4), mean over migrations and
seeds. Compilation time is excluded.}
\label{tab:cost}
\resizebox{\columnwidth}{!}{\begin{tabular}{lrrrrrr}
\toprule
 & \multicolumn{3}{c}{Development} & \multicolumn{3}{c}{Test} \\
\cmidrule(lr){2-4}\cmidrule(lr){5-7}
Model & Calls & Seconds & Repairs & Calls & Seconds & Repairs \\
\midrule
Qwen2.5-Coder 7B & 13.2 & 21 & 0.4 & 9.7 & 15 & 0.5 \\
Llama 3.1 8B & 16.3 & 31 & 0.1 & 9.7 & 23 & 0.0 \\
Gemma 3 4B & 17.4 & 30 & 0.0 & 9.7 & 16 & 0.3 \\
\bottomrule
\end{tabular}
}
\end{table}

\paragraph{Cost.} Table~\ref{tab:cost} reports generation cost. A migration needs ten to twenty model calls and well
under a minute of generation (\testCostQwenCoderCalls{} calls and \testCostQwenCoderSeconds{}\,s for
Qwen2.5-Coder on the test set). Stage~4 additionally configures a kernel tree at each training commit it checks
and compiles single object files there; build results are cached by commit, file and content.

\section{Threats to Validity}
\label{sec:threats}

\paragraph{What success measures.} Success requires the rule's output to match the human fix (exactly or as a
complete subset) and to compile. It does not test behaviour: a compiled conversion can still be wrong at run time,
and we run no driver tests. It also undercounts: a correct rule that converts a site the human converted in a later
commit, or writes an equivalent expression, is counted as a failure, and \emph{divergent} outputs
(Table~\ref{tab:safety}) include such cases as well as real errors. We did not label them manually; the released
outputs and a blind labelling packet (source hidden) allow this check. The undercount affects all systems, but not
necessarily equally.

\paragraph{Compilation scope.} Outputs are compiled for x86\_64 \texttt{allmodconfig} only. Targets whose original
file does not compile there (other architectures, broken intermediate commits) are excluded from the denominator
for every system alike. A rule that breaks a configuration we do not build would go unnoticed.

\paragraph{Contamination.} The local models' training cutoffs are taken from their vendors' statements. The test
set admits only migrations whose new API first appears in the kernel after all three cutoffs, but a model may still
have seen similar older APIs. The frontier-LLM baseline has no such guarantee and, unlike the local models, worked
agentically with a test loop of its own design; it is a reference point, not a like-for-like competitor.

\paragraph{Test-set construction.} Conversions are mined with line patterns, so the data contain only conversions
whose diff matches them, and precision was verified on a sample (at most eight per migration). The verification
was planned for one reviewer; we used two independent LLM reviewers instead (a judge and a skeptic) and report this
deviation from the pre-registered protocol. They disagreed on no sample's verdict. The frontier baseline's rule
writers were audited for file access; one agent saw the names (not contents) of another migration's training
folder through a reused scratch script, which we report in the released audit notes.

\paragraph{Development reuse.} Prompts, hints, thresholds and both critics were designed while looking at the
development migrations and their held-out results. Development numbers are therefore optimistic; only the test-set
numbers, obtained once after freezing the framework at a git tag, measure generalisation. The test-set
generation run was interrupted during its second configuration, first by the host running out of memory and
then when we moved it into a detached container (no output had been evaluated); the interrupted configuration was
restarted from scratch for the test migrations with the same seeds, and no completed
configuration was rerun.

\paragraph{Scale.} The test set has \testMigrations{} migrations and \testBuildable{} buildable targets, and three
seeds per configuration. Per-migration results swing with whether one leaf rule is right, so we report seed ranges
and per-migration tables rather than significance tests. All timings come from one laptop GPU and 4-bit models.

\section{Conclusion}
\label{sec:conclusion}

Earlier hand conversions of a kernel API migration are enough material for a 7B code model on a laptop to write
working Coccinelle rules, provided the work is structured for it: decomposed by variant, guided by the examples'
minimal edits, checked against the examples, repaired where the errors are mechanical, and gated by the compiler.
Prompted directly, the same model solves almost nothing on migrations it cannot have seen; with the framework it
approaches Spinfer and, on the test set, is more precise than every other system we tested. The structure does not rescue
general-purpose 4--8B models, and no local model handled changes that couple a declaration with its uses.

Three directions follow. Rules that must agree across a declaration and its call sites need an interface the model
cannot get wrong, for example a framework-written skeleton that fixes the shared metavariables. A semantic check
beyond compilation would catch outputs such as \texttt{sizeof} of a pointer. And a larger pre-registered test set,
refreshed as new APIs land, would let such frameworks be compared without contamination.

\section*{Use of AI Assistants}
The frontier-LLM baseline is a Claude agent (Anthropic), as described in Section~\ref{sec:setup}. The author also
used Claude as a programming and writing assistant for the experiment scripts and the text of this paper. Every
number in the paper is generated by the released scripts from the released result files, and the author takes
full responsibility for the content.

\bibliographystyle{IEEEtran}
\bibliography{refs}

@inproceedings{padioleau2006understanding,
  author = {Padioleau, Yoann and Lawall, Julia L. and Muller, Gilles},
  title = {Understanding Collateral Evolution in {Linux} Device Drivers},
  booktitle = {Proceedings of the 2006 EuroSys Conference},
  year = {2006},
  doi = {10.1145/1217935.1217942}
}

@inproceedings{padioleau2008documenting,
  author = {Padioleau, Yoann and Lawall, Julia and Hansen, Ren{\'e} Rydhof and Muller, Gilles},
  title = {Documenting and Automating Collateral Evolutions in {Linux} Device Drivers},
  booktitle = {Proceedings of the 2008 EuroSys Conference},
  year = {2008},
  doi = {10.1145/1352592.1352618}
}

@inproceedings{andersen2008generic,
  author = {Andersen, Jesper and Lawall, Julia L.},
  title = {Generic Patch Inference},
  booktitle = {23rd IEEE/ACM International Conference on Automated Software Engineering (ASE)},
  year = {2008},
  doi = {10.1109/ASE.2008.44}
}

@inproceedings{serrano2020spinfer,
  author = {Serrano, Lucas and Nguyen, Van-Anh and Thung, Ferdian and Jiang, Lingxiao and Lo, David and Lawall, Julia and Muller, Gilles},
  title = {{SPINFER}: Inferring Semantic Patches for the {Linux} Kernel},
  booktitle = {2020 USENIX Annual Technical Conference (USENIX ATC 20)},
  year = {2020}
}

@inproceedings{lawall2017prequel,
  author = {Lawall, Julia and Palinski, Derek and Gnirke, Lukas and Muller, Gilles},
  title = {Fast and Precise Retrieval of Forward and Back Porting Information for {Linux} Device Drivers},
  booktitle = {2017 USENIX Annual Technical Conference (USENIX ATC 17)},
  year = {2017}
}

@inproceedings{shariffdeen2021fixmorph,
  author = {Shariffdeen, Ridwan and Gao, Xiang and Duck, Gregory J. and Tan, Shin Hwei and Lawall, Julia and Roychoudhury, Abhik},
  title = {Automated Patch Backporting in {Linux} (Experience Paper)},
  booktitle = {Proceedings of the 30th ACM SIGSOFT International Symposium on Software Testing and Analysis (ISSTA)},
  year = {2021},
  doi = {10.1145/3460319.3464821}
}

@misc{yusuf2025howfar,
  author = {Yusuf, Imam Nur Bani and Huang, Xiping and Lawall, Julia and Jiang, Lingxiao and Lo, David and Bui, Nghi D. Q. and Li, Chunmiao and Wu, Haitao and Yu, Yijun},
  title = {Deep Learning-Based Automatic Semantic Patch Generation for the {Linux} Kernel: How Far Are We?},
  note = {Preprint, \url{https://imamnurby.github.io/assets/files/spatch_emse.pdf}},
}

@misc{kharlamova2025autodriver,
  title = {{LLM}-Driven Kernel Evolution: Automating Driver Updates in {Linux}},
  author = {Kharlamova, Arina and Liu, Jiawen and Zhang, Tianyi and Yang, Xinrui and Alqasimi, Humaid and Sun, Youcheng and Xue, Chun Jason},
  year = {2025},
  eprint = {2511.18924},
  note = {arXiv:2511.18924},
  archivePrefix = {arXiv}
}

@inproceedings{dang2025miggpt,
  title = {{MigGPT}: Harnessing Large Language Models for Automated Migration of Out-of-Tree {Linux} Kernel Patches Across Versions},
  author = {Dang, Pucheng and Huang, Di and Li, Dong and Chen, Kang and Wen, Yuanbo and Guo, Qi and Hu, Xing},
  booktitle = {Advances in Neural Information Processing Systems (NeurIPS)},
  year = {2025},
  note = {arXiv:2504.09474}
}

@inproceedings{li2026portgpt,
  title = {{PortGPT}: Towards Automated Backporting Using Large Language Models},
  author = {Li, Zhaoyang and Yu, Zheng and Song, Jingyi and Xu, Meng and Luo, Yuxuan and Mu, Dongliang},
  booktitle = {IEEE Symposium on Security and Privacy (S\&P)},
  year = {2026},
  note = {arXiv:2510.22396}
}

@inproceedings{almorsi2025gcg,
  title = {Guided Code Generation with {LLMs}: A Multi-Agent Framework for Complex Code Tasks},
  author = {Almorsi, Amr and Ahmed, Mohanned and Gomaa, Walid},
  year = {2024},
  booktitle = {Proceedings of the 2024 IEEE International Japan-Africa Conference on Electronics, Communications and Computations (JAC-ECC)},
  eprint = {2501.06625},
  note = {arXiv:2501.06625},
  archivePrefix = {arXiv}
}

@misc{qwen25coder,
  title = {Qwen2.5-Coder Technical Report},
  author = {Hui, Binyuan and Yang, Jian and Cui, Zeyu and others},
  year = {2024},
  eprint = {2409.12186},
  note = {arXiv:2409.12186},
  archivePrefix = {arXiv}
}

@misc{llama3,
  title = {The {Llama} 3 Herd of Models},
  author = {{Llama Team, AI @ Meta}},
  year = {2024},
  eprint = {2407.21783},
  note = {arXiv:2407.21783},
  archivePrefix = {arXiv}
}

@misc{gemma3,
  title = {{Gemma} 3 Technical Report},
  author = {{Gemma Team} and Kamath, Aishwarya and others},
  year = {2025},
  eprint = {2503.19786},
  note = {arXiv:2503.19786},
  archivePrefix = {arXiv}
}

@misc{ahmed2026siblings,
  author = {Ahmed, Mohanned},
  title = {Sibling Migrations as Free Supervision for {Linux} Driver Updates: A Pilot Study},
  year = {2026},
  howpublished = {\url{https://github.com/MohannedAhmed67/kernel-sibling-migrations}},
  note = {Code and data; see RESULTS.md}
}
\end{document}